\documentclass[11pt,psfig]{article}
\usepackage{graphicx}
\usepackage{bm}
\usepackage[centertags]{amsmath}
\usepackage{amsfonts}
\usepackage{amssymb}
\usepackage{amsthm}
\usepackage{newlfont}
\newlength{\defbaselineskip}
\newcommand{\onehalfspacing}{\setlength{\baselineskip}{1.5 \defbaselineskip}}

\newcommand{\be}{\begin{equation}}
\newcommand{\ee}{\end{equation}}
\newcommand{\bea}{\begin{eqnarray}}
\newcommand{\eea}{\end{eqnarray}}
\newcommand{\nno}{\nonumber}

\newcommand{\vp}{\varphi}
\newcommand{\vth}{\vartheta}

\begin{document}

\begin{flushright}
hep-th/0409143
\end{flushright}

\begin{center}
{\Large\bf Spinning Test Particle in Kalb-Ramond background}
\\[12mm]
Debaprasad Maity \footnote{E-mail: tpdm@iacs.res.in}, ~~Soumitra SenGupta \footnote{E-mail: 
tpssg@iacs.res.in}, ~ \\
{\em Department of Theoretical Physics\\ Indian Association for the
Cultivation of Science\\ Calcutta - 700 032, India} \\
\bigskip

Sourav Sur \footnote{E-mail: sourav@iopb.res.in}\\
{\em Institute of Physics, Bhubaneswar - 751005, 
India}\\[10mm]
\end{center}

\begin{abstract}

In this work we explore the geodesic deviations of spinning test particles in a string inspired
Einstein-Kalb Ramond background. Such a background is known to be equivalent to
a spacetime geometry with torsion. We have shown here that the antisymmetric Kalb-Ramond field has 
significant effect on the geodesic deviation of a spinning test particle.
A search for an observational evidence of such an effect in astrophysical experiments may lead to
a better undestanding of the geometry of the background spacetime.
\end{abstract}

\onehalfspacing

\section{Introduction       \label{gr-intro}}

String theory, since it's emergence, is considered to be the most promising candidate
for a consistent perturbative quantum theory of gravity. The search for the signature
of string theory in the low energy world has intensified over last few years to establish
a contact of this theory with the real world. The possible testing grounds are
the accelerator experiments and cosmological/astrophysical observations. The present
work aims to search for such a stringy signal in an astrophysical observation through the study
of geodesic deviations of spinning test particles. It is well known that the low energy limit
of the gravity sector of string theory indeed reproduces the curved spacetime picture as
proposed in Einstein's theory. However the presence of the massless second rank antisymmetric
tensor field (Kalb-Ramond field) \cite{kr} endows the background spacetime with torsion. Thus a
string inspired background differs from Einstein's framework by a Cartan extension.

In Einstien's framework the dynamics of particles in curved spacetime has been an important
subject of investigation. Since the early stages of the development of general relativity, the study
of the motion of a test particle (i.e., a particle which is sufficiently small compared to other objects
producing the field and has a negligible influence on the field) in a curved background is of great
importance. Studies of the dynamics of streams of cosmic particles in the astrophysical/cosmological
experiments reveal the important properties of the background spacetime. As is well-known a test particle
of the simplest type, i.e., one without any internal structure, has been shown to follow the so-called 
geodesics.
Such test particles with single pole structures are referred to as `pole particles'. However, a 
test particle can
have a structure of its own, thereby giving rise to a non-vanishing spin-density for the particle. 
As such, its
equation of motion can then depend on this structure. A test particle with such a {\it multipole} 
structure is
expected to follow a trajectory that is different from that of the usual geodesic. A theory describing 
the motion
of such `pole-multipole particles' has been developed initially by Papapetrou \cite{ppt} and later on by 
Dixon
\cite{dixon} in an alternative approach. Subsequent applications of the Papapetrou formalism to the 
particular
case of motion in the static spherically symmetric Schwarzschild field has been carried out by 
Corinaldesi and
Papapetrou \cite{conppt} and since then there has been a growing interest in the study of motion of 
spinning
test particles under the influence of gravity. Most recently, there had been a plethora of works in the 
literature
\cite{spinref} that investigate the dynamics of spinning test particles in different sorts of 
background spacetimes.

In this work, we study the motion of spinning particles in a general static spherisymmetric spacetime in 
presence
of the Kalb-Ramond (KR) field. It has been shown \cite{pmssg} in the context of string theory that the 
KR field, in
general, has a gauge invariant coupling with the electromagnetic field, and produces significant 
effects on many
cosmological/astrophysical phenomena \cite{skpmssgas,skpmssgss,skssgss,ssgssEL}. 
Extensive works have
also been carried out \cite{bmsenssg,bmsensenssg} to explore the role of such antisymmetric tensor 
background
in the context of compact extra dimensional theories of Arkani-Hamed--Dimopoulous--Dvali (ADD) \cite{add} 
and
Randall-Sundrum (RS) \cite{rs} types. The observational possibilities of a stringy signal has also been 
discussed
in several other works \cite{dmssg,dmpmssg}. Here we explore  another possible experimental signature 
of such
a string inspired background through the influence of KR field on the geodesics of spinning 
test particles. We
show that such geodesics indeed differs from those observed in a pure Einstein's background. 
An estimate of this departure is made in terms of the KR field and spin of the test particle.

\vskip .4in
\section{General equation of motion of spinning test particle  \label{gr-eom}}

According to the formalism of Papapetrou \cite{ppt}, the trajectory of a spinning
test particle in an arbitrary spacetime structure is shown to deviate from the
usual geodesic and is described by
\be
\frac D {D \tau} \left(m ~u^{\alpha} ~+~ u_{\beta} \frac { DS^{\alpha\beta}}
{D \tau}\right) ~+~ \frac 1 2~ R^{\alpha}_{\mu\nu\lambda}~ u^{\nu}~
S^{\lambda\mu} ~=~ 0           \label{gr-spintraj}
\ee
where $\tau$ is the proper time, $m$ and $u^{\alpha}$ are respectively the
particle's mass and four-velocity, $S^{\alpha \beta}$ is the antisymmetric
spin tensor of the particle, $R^{\alpha}_{~ \mu \nu \lambda}$ is the curvature
tensor corresponding to the background field distribution on which the particle
moves, and $D/D\tau$ denotes a covariant derivative along $u^{\alpha}$ as
\be
\frac{D S^{\alpha \beta}}{D \tau} ~=~  \frac{d S^{\alpha\beta}}{d\tau} ~+~
\left(\Gamma^{\alpha}_{\mu\nu} ~S^{\mu\beta}  ~+~  \Gamma^{\beta}_{\mu\nu}~
S^{\alpha\mu}\right) u^{\nu}               \label{gr-spincov}
\ee
$\Gamma^{\alpha}_{\mu\nu}$ being the usual Christoffel connections.

The spin of the particle evolves as
\be
\frac {D S^{\alpha\beta}}{D \tau} ~+~ u^{\alpha} u_{\rho}~ \frac
{D S^{\beta\rho}}{D \tau} ~-~ u^{\beta} u_{\rho}~ \frac{D S^{\alpha\rho}}
{D \tau} ~=~ 0.        \label{gr-spin}
\ee
For vanishing spin one can easily verify that the above trajectory equation
(\ref{gr-spintraj}) reduces to the usual geodesic equation.

The above equations are, however, not sufficient to determine 
all the unknowns. It may be noted that  the number of independent equations to determine
of the spin components are three, while the number of independent spin components
are six. Therefore, to reduce the number of independent spin components one has
to impose a suitable supplementary condition that specifies the line $L$ that
represents the motion of a `pole-dipole' particle inside the world tube of the
particle (see Papapetrou \cite{ppt}). The simplest supplementary condition
suggested by Corinaldesi and Papapetrou \cite{conppt} on investigating the
motion of a spinning test particle in a static spherisymmetric spacetime has been
\be
S^{0i}=0~;~~~ i = 1, 2, 3.       \label{gr-spinsup}
\ee
This condition, although not covariant, provides a very physical definition
of the representing world line $L$ of the spinning test particle. One can, in
fact, show that if the coordinates of $L$ are designated by $X^{\alpha}$ (which
are functions of the proper time $\tau$ along $L$) then in the rest frame of the
central attracting body each point $X \in L$ coincides with the center of mass of
the particle. Other kinds of supplementary conditions can also be found in the
literature \cite{spinsupp}, however we presently resort to the above condition
owing to the physical relevance mentioned above.

Considering a general static spherically symmetric spacetime metric structure, viz.,
\be
d\tau^2 ~=~ - e^{\nu (r)} dt^2 ~+~ e^{\lambda (r)} dr^2 ~+~ r^2 \left(d \vth^2
~+~ \sin^2 \vth d \vp^2\right)     \label{gr-metric}
\ee
one can obtain, on using the above supplementary equation (\ref{gr-spinsup}), the
explicit form of the spin evolution equations (\ref{gr-spin})
\bea
&(a)&~~~ \dot{S}^{12} ~+~ \left(\frac{\lambda'} 2 ~+~ \frac 1 r ~-~ \frac{\nu'}
2\right) \dot{r}~ S^{12} ~+~ r e^{-\lambda}~ \sin^2 \vth~ \dot{\vp}~ S^{23} ~+~
\cos \vth~ \sin \vth ~\dot{\vp}~ S^{31} ~=~ 0 \nno\\
&(b)&~~~ \dot{S}^{23} ~+~ \left(\frac{\nu'} 2 ~-~ \frac 1 2\right) \dot{\vp}~
S^{12} ~~+~ \left(\frac {2 \dot{r}} r ~+~ \cot \vth~ \dot{\vth}\right) S^{23}
~+~ \left(\frac{\nu'} 2 ~-~ \frac 1 r\right) \dot{\vth}~ S^{31}
~~=~~ 0 \nno\\
&(c)&~~~ \dot{S}^{31} ~-~ \cot \vth~ \dot{\vp}~ S^{12} ~+~ r e^{-\lambda}
~\dot{\vth}~ S^{23} ~+~ \left[\left(\frac 1 r ~+~ \frac{\lambda'} 2 ~-~
\frac{\nu'} 2\right) \dot{r} ~+~ \cot \vth~ \dot{\vth}\right] S^{31}
=~ 0 .           \label{gr-spinevol}
\eea
where the overhead dot indicates differentiation with respect to $\tau$ and a
prime denotes differentiation with respect to $r$.  The equations of motion
(\ref{gr-spintraj}) for the test particle now take the form
\bea
&(a)&~~~ \frac d {d \tau} \left[\left(m ~+~ m_s\right) \dot{t}\right] ~+~
\left(m ~+~ m_s\right) \Gamma^0 ~=~ 0  \nno\\
&(b)&~~~ \frac d {d \tau} \left[\left(m ~+~ m_s\right) \dot{r}\right] ~+~
\left(m ~+~ m_s\right) \Gamma^1 \nno\\
&& \hskip 2in ~+~ r e^{-\lambda} \left(\frac{\lambda'} 2
~+~ \frac{\nu'} 2\right) \left(S^{12}~ \dot{\vth} ~-~ S^{31}~ \sin^2 \vth~
\dot{\vp}\right) ~=~ 0  \nno\\
&(b)&~~~ \frac d {d \tau} \left[\left(m ~+~ m_s\right) \dot{\vth}\right] ~+~
\left(m ~+~ m_s\right) \Gamma^2 ~+~ r \left(\frac{\lambda' \nu'} 4 ~-~
\frac{\nu'^2} 4 ~-~ \frac{\nu''} 2 ~-~ \frac{\lambda'} {2 r}\right) S^{12} \nno\\
&&\hskip 2.5in +~ r e^{-\lambda}~\sin^2 \vth~ \dot{\vp} \left(\frac{\nu'} 2 ~+~
\frac{e^{\lambda}} r ~-~ \frac 1 r\right) S^{23} ~=~ 0 \nno\\
&(d)&~~~ \frac d {d \tau} \left[\left(m ~+~ m_s\right) \dot{\vp}\right] ~+~
\left(m ~+~ m_s\right) \Gamma^3 ~-~ r e^{-\lambda}~ \dot{\vth}
\left(\frac{\nu'} 2 ~+~ \frac{e^{\lambda}} r ~-~ \frac 1 r\right) S^{23} \nno\\
&&\hskip 2.5in -~ \dot{r} \left(\frac{\lambda' \nu'} 4 ~-~ \frac{\nu'^2} 4 ~-~
\frac{\nu''} 2 ~-~ \frac{\lambda'}{2r}\right) S^{31} ~~=~~ 0  \label{gr-spineom}
\eea
where ~~~$\Gamma^{\mu} ~\equiv~ \Gamma^{\mu}_{\nu\lambda}~ u^{\nu}~ u^{\lambda}$~~~
with components
\bea
\Gamma^{0} &=& \nu'~ \dot{r}~ \dot{t} \nonumber\\
\Gamma^{1} &=& \frac{\lambda'} 2~ \dot{r}^2 ~-~ r e^{-\lambda}~ \dot{\vth}^2
~-~ r e^{-\lambda} ~\sin^2 \vth ~\dot{\vp}^2 ~+~ e^{(\nu ~-~ \lambda)} ~\nu'
~\dot{t}^2 \nonumber\\
\Gamma^{2} &=& \frac 2 r~ \dot{r}~ \dot{\vth} ~-~ \sin \vth~ \cos \vth~
\dot{\vp}^2 \nonumber\\
\Gamma^{3} &=& \frac 2 r~ \dot{r}~ \dot{\vp} ~+~ 2~ \cot \vth~ \dot{\vth}~
\dot{\vp}.       \label{gr-spingamma}
\eea
Here $m$ is the particle's mass and the quantity $m_s$ is defined by
\be
m_s ~=~ \frac{r^2} 2~ \nu'~ \left(\sin^2 \vth ~\dot{\vp} ~S^{31} ~-~
\dot{\vth}~ S^{12}\right) .        \label{gr-spinmass}
\ee
$m_s$ can be viewed as an effective mass originating from the spin-orbit coupling
and $(m + m_s)$ is the total effective mass.

In what follows, we shall be investigating whether the above two sets of
equations (\ref{gr-spinevol}) and (\ref{gr-spineom}) admit any solution
representing a motion on a plane passing through the central body,
of course, with the view that both the mass and the spin of the
test particle have exceedingly small effects on the background
spacetime. As usual, one can take without any loss of generality,
this plane as the equatorial plane ~$\vth = \pi/2$~ making the
simple choice
\be
S^{31} ~\neq 0~;~~~~~~ S^{12} ~=~ S^{23} ~=~ 0          \label{gr-spinchoice}
\ee
whence ~$m_s ~=~ (r^2/2)~ \nu'~ \sin^2 \vth~ \dot{\vp}~ S^{31}$~.
This leads to the following set of equations of motion, 
\bea
&(a)&~~~ \dot{S}^{31} ~+~ \left(\frac 1 r - \frac{\nu' ~-~ \lambda'} 2\right)
\dot{r}~ S^{31} ~=~ 0  \nno\\
&(b)&~~~ \frac d {d \tau} \left[\left(m ~+~ m_s\right) \dot{t}\right] ~+~
\left(m ~+~ m_s\right) \nu'~ \dot{r}~ \dot{t} ~=~ 0  \nno\\
&(c)&~~~ \frac d {d \tau} \left[\left(m ~+~ m_s\right) \dot{\vp}\right] ~+~
\left(m ~+~ m_s\right) \frac{2 \dot{r} \dot{\vp}} r ~+~ \left(\frac {\nu''} 2
~+~ \frac {\nu'^2} 4 ~-~ \frac{\nu' \lambda'} 4 ~+~ \frac{\lambda'}{2 r}\right)
\dot{r}~ S^{31} ~=~ 0 \nno\\
&(d)&~~~ \frac d {d \tau} \left[\left(m ~+~ m_s\right) \dot{r}\right] ~+~
\left(m ~+~ m_s\right) \left\{\frac{\dot{r}^2 \lambda'} 2 ~-~ r e^{-\lambda}
~\dot{\vp}^2 ~+~ \frac{\nu' \dot{t}^2} 2~ e^{(\nu ~-~ \lambda)}\right\} \nno\\
&& \hskip 3in -~ r e^{-\lambda}~ \left(\frac{\nu' ~+~ \lambda'} 2\right)~
\dot{\vp}~ S^{31} = 0 .            \label{gr-spinsimp}
\eea
Eq.(\ref{gr-spinsimp} b) at once gives the integral of energy
\be
e^{\nu} ~\dot{t} \left(m ~+~ m_s\right) ~=~ E ~(\hbox{const.})  \label{gr-spinenergy}
\ee
while Eq.(\ref{gr-spinsimp} a) stands as a first integral of spin only
\be
r~ S^{31}~ e^{(\lambda ~-~ \nu)/2} ~=~ K ~(\hbox{const.}) .     \label{gr-spinspin}
\ee
Using Eq.(\ref{gr-spinenergy}) and Eq.(\ref{gr-spinspin})
and noticing that
\be
\frac{d^2}{d\tau^2} ~=~ \dot{t}^2~ \frac{d^2}{dt^2} ~+~ \ddot{t}~
\frac d {dt}~ \nno
\ee
the variable $\tau$ can be eliminated from the equations (\ref{gr-spinsimp} c)
and (\ref{gr-spinsimp} d), whence we obtain
\bea
&(a)&~~~ \frac {d^2 r}{dt^2} ~+~ \left(\frac{\lambda'} 2 - \nu'\right)
\left(\frac{dr}{dt}\right)^2 ~-~ r e^{- \lambda} \left(\frac{d \vp}{dt}\right)^2
~+~ \frac{\nu'} 2 ~e^{(\nu - \lambda)} ~-~ \frac K E ~e^{\frac{3 (\nu - \lambda)} 2}~
\frac{(\nu' +\lambda')} 2 ~\frac {d \vp}{dt} ~= 0 \nno\\
&(b)&~~~ \frac {d^2 \vp}{dt^2} ~+~ \left(\frac 2 r - \nu'\right) \frac{dr}{dt}~
\frac{d \vp}{dt} ~+~ \frac K E \left(\frac {\nu''} 2
~+~ \frac {\nu'^2} 4 ~-~ \frac{\nu' \lambda'} 4 ~+~ \frac{\lambda'}{2 r}\right)
e^{\frac{(3 \nu - \lambda)} 2} ~\frac 1 r~ \frac{dr}{dt} ~=~ 0 .  \label{gr-geneom}
\eea

\vskip .4in
\section{Spinning test particle trajectory in static spherisymmetric
Einstein--Kalb-Ramond spacetime      \label{gr-traj}}

Following the formalism in \cite{pmssg} the solutions for the metric coefficients
in a general static spherical symmetric spacetime involving the KR field has been
obtained in \cite{ssgss,skssgss}
\bea
e^{\nu (r)} &=& 1 ~-~ \frac{r_s} r ~+~ b \left[ \frac {r_s}{6 r^3}
+ \frac{r_s^2}{6 r^4} ~+~ \frac{3 r_s (r_s^2 ~-~ b/2)}{20 r^5} ~+~ \cdots
\right] \nonumber\\
e^{- \lambda (r)} &=& 1 ~-~ \frac{r_s} r ~+~ b \left[ \frac 1 {r^2}
~+~ \frac {r_s}{2 r^3} ~+~ \frac{r_s^2}{3 r^4} ~+~ \frac{r_s (r_s^2 ~-~ b/6)}
{4 r^5} ~+~ \cdots \right].  \label{gr-krmetric}
\eea
where $r_s = 2 G M$ is the Schwarzschild radius and the constant $b$ is the
measure of the strength of KR field (which has a natural interpretation in the
form of a background torsion). The parameter $b$ can be negative or positive
depending the nature of the torsion--KR field coupling constant within a minimal
coupling prescription as has been mentioned in \cite{ssgssJCAP}. Accordingly,
torsion may exhibit a repulsive (anti-gravitating) or an attractive character. In the
special case of vanishingly small gravitating mass $M = 0$, i.e., $r_s = 0$, the above
solutions reduce to the simple closed-form structures \cite{ssgss}
\be
e^{\nu (r)} ~=~ 1~;~~~~~  e^{- \lambda (r)} ~=~ 1 ~-~
\frac b {r^2}. \label{gr-krmetricapp}
\ee
Depending on positive or negative value of $b$, these represent a `wormhole' of
throat radius $\sqrt{|b|}$ or a `naked singularity' at $r = 0$.

We first study the dynamics of the spinning test particle in the special case $r_s
= 0$ and then follow up with a more rigorous investigation in the general case
$r_s \neq 0$.

\vskip .3in
\subsection{An otherwise empty spacetime in presence of Kalb-Ramond
field  \label{gr-empty}}

In the case $(r_s = 0)$, when the Einstein-KR spacetime is otherwise empty, the
equations of motion \ref{gr-geneom} for the spinning test particle reduce to
\bea
&(a)&~~~ \frac{d^2 r}{dt^2} ~+~ \frac{\lambda'} 2 \left(\frac{dr}{dt}\right)^2 -~
r e^{-\lambda} \left(\frac{d\vp}{dt}\right)^2 -~ \frac K E~ \frac{\lambda'} 2~
e^{- 3 \lambda/2}~ \frac{d \vp}{dt} ~=~ 0 \nno\\
&(b)&~~~ \frac {d^2 \vp}{dt^2} ~+~ \frac 2 r~ \frac{dr}{dt}~ \frac{d \vp}{dt} ~+~
\frac K E~ \frac{\lambda'}{2~ r^2} ~e^{- \lambda/2}~ \frac{dr}{dt}
~=~ 0 .      \label{gr-kreom}
\eea
Eq.(\ref{gr-kreom} b) for $\vp$ yields the integral of angular momentum
\be
r^2~ \frac{d \vp}{dt} ~-~ \frac K E~ e^{- \lambda/2} ~=~
I_s ~(\hbox{const.}) .      \label{gr-eangmom}
\ee
The equation (\ref{gr-kreom} a) for $r$ then reduces to
\be
\frac{d \vp}{dU} ~=~ \frac{e^{\lambda/2} ~+~ \sigma}
{\sqrt{(1 ~+~ \sigma~ e^{- \lambda_0/2})^2 ~-~ U^2~ (1 ~+~
\sigma~ e^{- \lambda/2})^2}}        \label{gr-kreomf}
\ee
where we have used the dimensionless independent variable ~$U = r_0/r$ ,~$r_0$
being the distance of closest approach of the spinning test particle towards
the center of force; $\lambda_0 \equiv \lambda(r_0)$~ and as such ~$e^{-\lambda (U)}
= 1 - \mu U^2$, where $\mu = b/r_0^2$ is the dimensionally scaled KR parameter.
$\sigma = K/(E I_s)$ is a dimensionless parameter that contains the integrals
of energy, spin and orbital angular momentum which describe the trajectory of
the test particle.
\begin{figure}
\includegraphics[width=7in,height=3.25in]{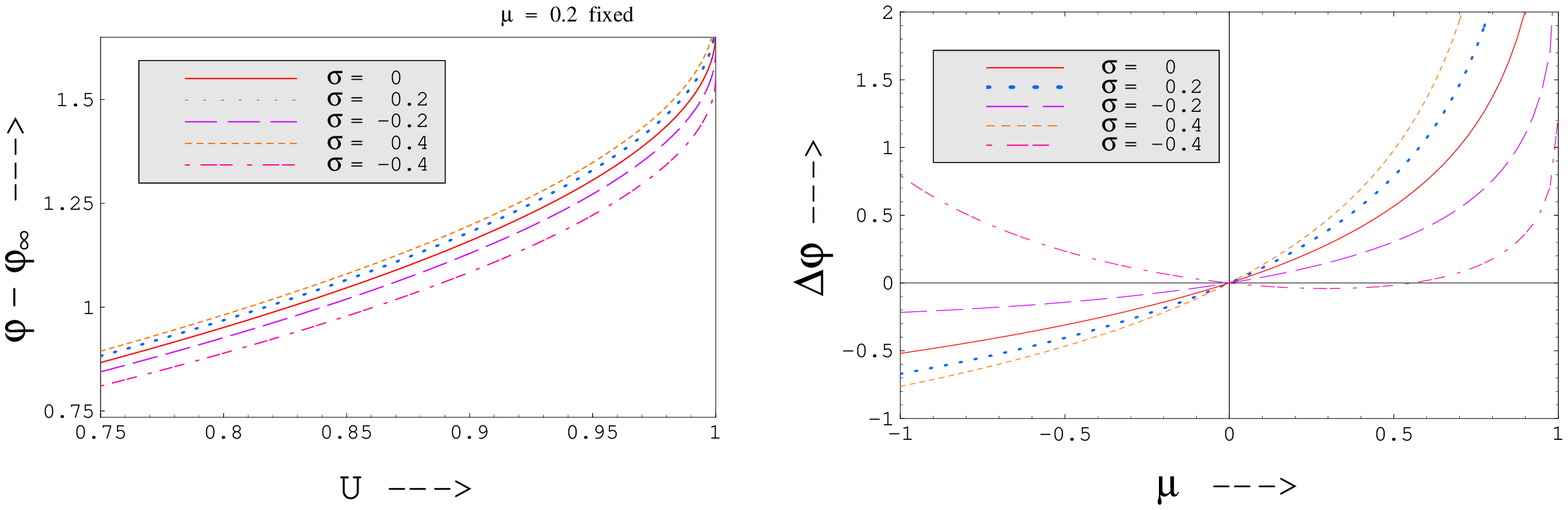}
{\small .}\hskip 1.9in (a) \hskip 3in (b)
\vskip .1in
\caption{ (a) Variation of $[\vp (r) - \vp_{\infty}]$ with $U = r_0/r$ for a
characteristically chosen fixed value $(= 0.2)$ of the parameter $\mu = b/r_0^2$
and a range of values of the parameter $\sigma$ from $- 0.4$ to $0.4$. In order
to achieve appreciable deviations the plots have been exaggerated for higher values
of $U$ (close to unity). \hskip 1in
(b) plots of the angle of bending $\Delta \vp$ as a function of $\mu$ for
parametric values of $\sigma$ ranging from $- 0.4$ to $0.4$.}    \label{gr-spines}
\vskip .2in
\includegraphics[width=7in,height=3.25in]{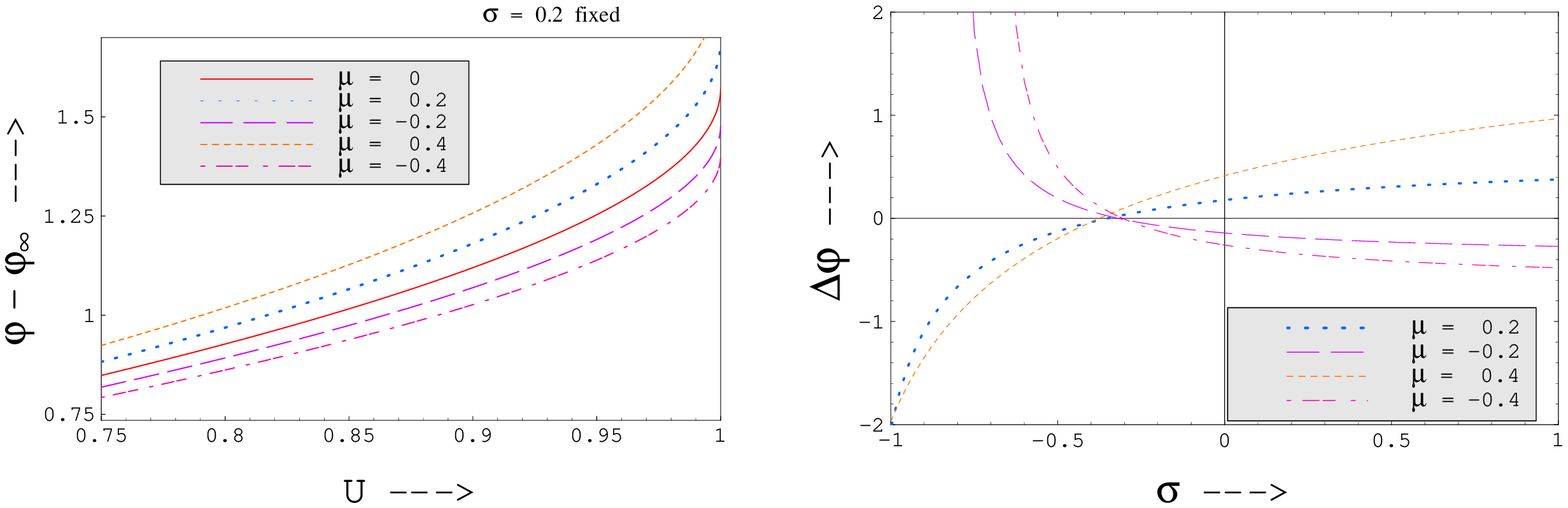}
{\small .}\hskip 1.9in (a) \hskip 3in (b)
\vskip .1in
\caption{ (a) Variation of $[\vp (r) - \vp_{\infty}]$ with $U = r_0/r$ for a
characteristically chosen fixed value $(= 0.2)$ of the parameter $\sigma$
and a range of values of the parameter $\mu$ from $- 0.4$ to $0.4$. In order
to achieve appreciable deviations the plots have been exaggerated for higher values
of $U$ (close to unity). \hskip 1.1in
(b) plots of the angle of bending $\Delta \vp$ as a function of $\sigma$ for
parametric values of $\sigma$ ranging from $- 0.4$ to $0.4$. For $\mu = 0$, however,
$\Delta \vp = 0$.}    \label{gr-spineb}
\end{figure}

As we are primarily interested in obtaining the equation of trajectory of the
spinning test particle at distances far away from the center of force ($r \gg r_0$ ,
i.e., $U \ll 1$) where the KR field is sufficiently weak ($|b| \ll r^2$ for all
$r \geq r_0$), we make a large-distance approximation upto order $U^2$ and write
the above equation in the integral form as
\be
\int d\phi ~\approx~ \frac 1 {1 ~+~ \sigma} ~\int \frac{du}{\sqrt{\delta^2 ~-~ U^2}}
\left\{1 ~+~ \sigma ~+~ \frac {\mu U^2} 2 ~+~ {\cal O}\left(\mu U^2\right)^2\right\}
~;~~~~~~~U ~=~ \frac {r_0} r \ll 1      \label{gr-spinapp}
\ee
where
\be
\delta ~=~ \frac 1 {1 ~+~ \sigma} \left[1 ~+~ \sigma
\left\{1 ~-~ \frac {\mu} 2 ~+~ {\cal O}\left(\mu^2\right)
\right\}\right]~;~~~~~~~~~~ |\mu| ~=~ \frac {| b |}{r_0^2} \ll 1 .
\ee
The solution of Eq.(\ref{gr-spinapp}) is given as
\be
\vp ~-~ \vp_\infty ~=~ \sin^{-1} \frac U {\delta} ~+~ \frac {\mu~
\delta^2}{4~ (1 ~+~ \sigma)} \left[\sin^{-1} \frac U {\delta} ~-~
\frac U {\delta}~ \sqrt{\delta^2 ~-~ U^2}\right] ~+~ {\cal O}
(\mu U^2)^2 .        \label{gr-spinappsol}
\ee
where $\vp_\infty$ is the value of $\vp$ in the asymptotic limit $r \rightarrow
\infty$. The second term on the right hand side of Eq.(\ref{gr-spinappsol}) gives
the measure of bending of the trajectory of the spinning test particle (time-like
or null) in the Einstein--KR spacetime. This departure from the straight line motion 
which is a characteristic in Minkowski spacetime (in absence of the KR field) signals 
the exclusive effect of the antisymmetric background on the geodesic and may be called
as KR-lensing. The spin of the particle described by the parameter $\sigma$ does 
not contribute to the bending independently, but is coupled with the KR parameter
$\mu = b/r_0^2$ in an otherwise empty spacetime. In fact, the presence of the KR 
field makes the effect of the particle spin on trajectory deviation perceptable 
even when the gravitating mass is zero. The change inflicted by the spin parameter 
$\sigma$ on the expression for $(\vp ~-~ \vp_\infty)$ for a pure KR background (i.e., 
for $\sigma = 0$ --- the result in \cite{skssgss}) is given to the leading order in 
$\mu \sigma$ as 
$$ \frac {\sigma \mu} 4 \left[\left(2 ~-~ \sqrt{1 - U^2}\right) ~U ~-~ 
\sin^{- 1} U\right] .$$

Now, instead of performing an approximate integration of the equation of motion
(\ref{gr-kreomf}) for $U \ll 1$, one can also solve Eq.(\ref{gr-kreomf})
numerically, for characteristically chosen numerical values of the parameters
$\sigma$ and $\mu$. The amounts of trajectory deviation of the spinning test
particles in such a numerical evaluation are depicted in Fig. \ref{gr-spines}a~
and Fig. \ref{gr-spineb}a, while the variations of the angle of bending, viz.,~
$\Delta \vp = 2 \vert\vp - \vp_{\infty}\vert - \pi$~ as function of $\mu$ (for
fixed $\sigma$) or as function of $\sigma$ (for fixed $\mu$) are shown respectively
in Fig. \ref{gr-spines}b~ and Fig. \ref{gr-spineb}b.

While the above results are achieved in an idealized situation where a
vanishingly small gravitating mass $M$ (and hence $r_s$) is considered,
in the following section we investigate the motion of spinning test
particles in the more realistic scenario, that is of a general static
spherical spacetime background in presence of the KR field i.e with both $b$ and
$r_s \neq 0$.

\vskip .3in
\subsection{A general static spherically symmetric spacetime in
presence of Kalb-Ramond field  \label{gr-general}}

In the general static spherically symmetric spacetime $(r_s \neq 0)$ with the
metric in the form (\ref{gr-metric}),
we have the differential equations (\ref{gr-geneom} a \& b) for $r$ and $\vp$.
Eq.(\ref{gr-geneom} b) yields the general integral of angular momentum given by
\be
r^2 ~e^{-\nu} ~\frac {d \vp}{dt} - \frac K E~ e^{(\nu ~-~ \lambda)/2}
\left(1 ~-~ \frac{r \nu'} 2 \right) ~=~ I_g  ~(\hbox{const.}).   \label{gr-gangmom}
\ee
Considering the particle's spin to be small we obtain the equation of the orbit
in an power series expansion of a redefined dimensionless spin parameter $\sigma
= K/(E I_g)$ as
\be
\left(\frac{dr}{d \vp}\right)^2 =~ r^4~e^{- \lambda}~ \frac{\left[
\frac 1 {r_0^2} ~-~ \frac 1 {r^2} ~+~ \frac 1 {I_g^2} \left(e^{- \nu} ~-~
e^{- \nu_0}\right) ~+~ 2 \sigma \left(\frac 1 {r_0^2}~e^{(\nu_0 - \lambda_0)/2}
~-~ \frac 1 {r^2}~ e^{(\nu - \lambda)/2}\right) + {\cal O}\left(\sigma^2\right)\right]}
{\left[1 ~+~ \sigma \left(1 ~-~ \frac{r \nu'} 2\right) e^{(\nu ~-~ \lambda)/2}
\right]^2} .   \label{gr-spinorbit}
\ee
Due to extreme complexity of this equation in the general static spherisymmetric
Einstein-KR spacetime with solutions for $e^{\nu}$ and $e^{- \lambda}$ as
given in Eqs.(\ref{gr-krmetric}), we focus on the rather simplified scenario,
that of the dynamics of null (massless) spinning particles in such a spacetime.

Now, resorting to the limit $\sigma \rightarrow 0$, we have ~$I_g \rightarrow
I_0 = r^2 e^{- \nu} \dot{\vp}$~ and the above equation (\ref{gr-spinorbit}) can
be recast \cite{chandra,wein}
\be
\left(\frac{dr}{d \vp}\right)^2 =~ r^2~e^{- \lambda (r)}~ \left(\frac{r^2}{D^2}~
e^{- \nu (r)} ~-~ 1\right)      \label{gr-limspinorbit}
\ee
where $D = I_0$ is the impact parameter for null particles. The solution
is given in the form of a quadrature \cite{chandra,wein}
\be
\vp (r) ~-~ \vp_{\infty} ~=~ \int_r^{\infty} \frac{dr'}{r'} ~e^{\lambda (r')/2}~
\left(\frac{r'^2}{D^2}~e^{- \nu (r')} ~-~ 1\right)^{- 1/2} .     \label{gr-limsol}
\ee
At the distance of closest approach $(r_0)$ to the center of force, ~$dr/d\vp
\vert_{r=r_0} = 0$,~ whence Eq.(\ref{gr-limspinorbit}) gives
\be
D ~=~ I_0 ~=~ r_0~e^{- \nu_0/2} ,       \label{gr-I0}
\ee
where $\nu_0 \equiv \nu (r_0)$.

For non-vanishing $\sigma$, Eq.(\ref{gr-gangmom}) for the integral of angular
momentum can be rewritten as
\be
I_g ~=~ \frac{I_0}{1 ~+~ \sigma \left(1 ~-~ \frac{r \nu'}2\right) e^{(\nu ~-~
\lambda)/2}}~;~~~~~~~~~~ I_0 ~=~ r^2~ e^{- \nu}~ \dot{\vp}   \label{gr-Ig}
\ee
Using Eqs.(\ref{gr-I0}) and (\ref{gr-Ig}) and changing the independent variable
to $U \equiv r_0/r$, we finally write the equation for the orbit as
\be
\frac{d \vp}{dU} ~=~ \pm \frac{e^{\lambda/2}~ F (U)}{\sqrt{1 ~-~ U^2 ~+~
\left(e^{\nu_0 - \nu} ~-~ 1\right) F^2 (U) ~+~ 2 \sigma G (U) ~+~ {\cal O}
\left(\sigma^2\right)}}       \label{gr-fspinorbit}
\ee
where
\be
F (U) ~=~ 1 ~+~ \sigma \left(1 ~+~ \frac U 2~\frac{d \nu}{dU}\right) e^{(\nu ~-~
\lambda)/2}~; ~~~~~~~~
G (U) ~=~ e^{(\nu_0 ~-~ \lambda_0)/2} ~-~ U^2 e^{(\nu ~-~ \lambda)/2} . \label{gr-fg}
\ee
\begin{figure}
\includegraphics[width=7in,height=3.25in]{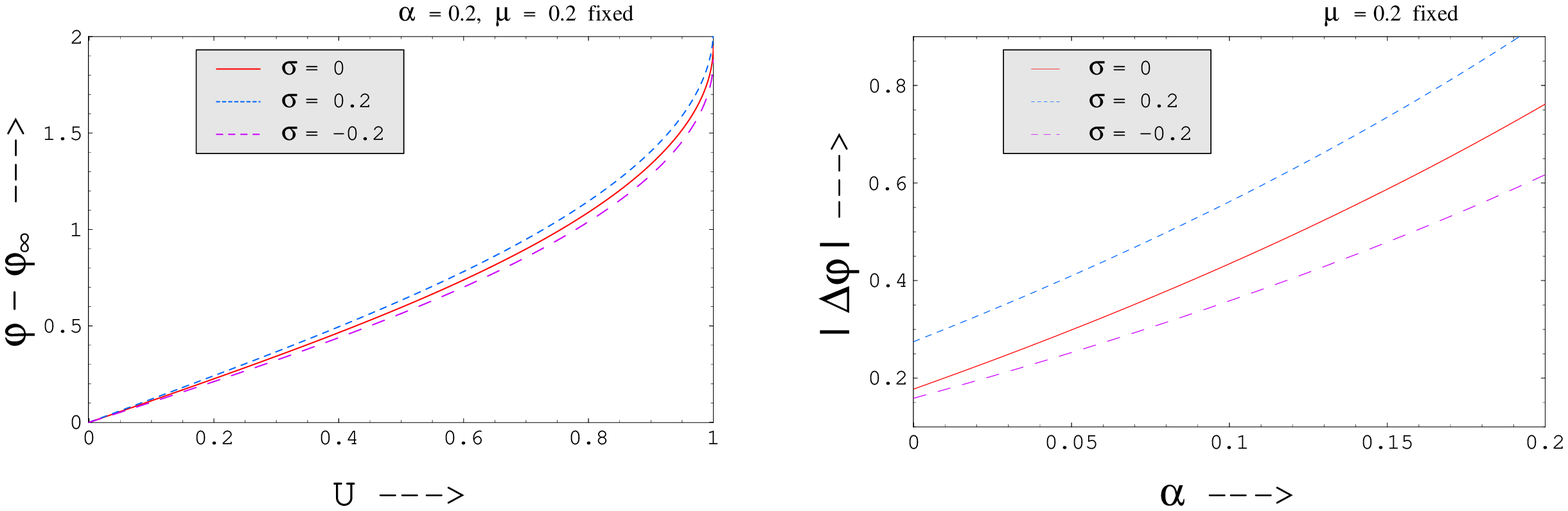}
{\small .}\hskip 1.9in (a) \hskip 3in (b)
\vskip .1in
\caption{ (a) Variation of $[\vp (r) - \vp_{\infty}]$ with $U = r_0/r$ for
characteristically chosen fixed values of the parameters $\alpha (= 0.2)$ and
$\mu (= 0.2)$ and three values $(= 0, 0.2$ and $- 0.2)$ of the parameter
$\sigma$. \hskip 1in
(b) plots of the magnitude of angle of bending $\Delta \vp$ as a function of $\alpha$ for a
fixed value of $\mu (= 0.2)$ and three parametric values $(= 0, 0.2$ and $- 0.2)$
of $\sigma$.}  \label{gr-spings}
\includegraphics[width=7in,height=3.25in]{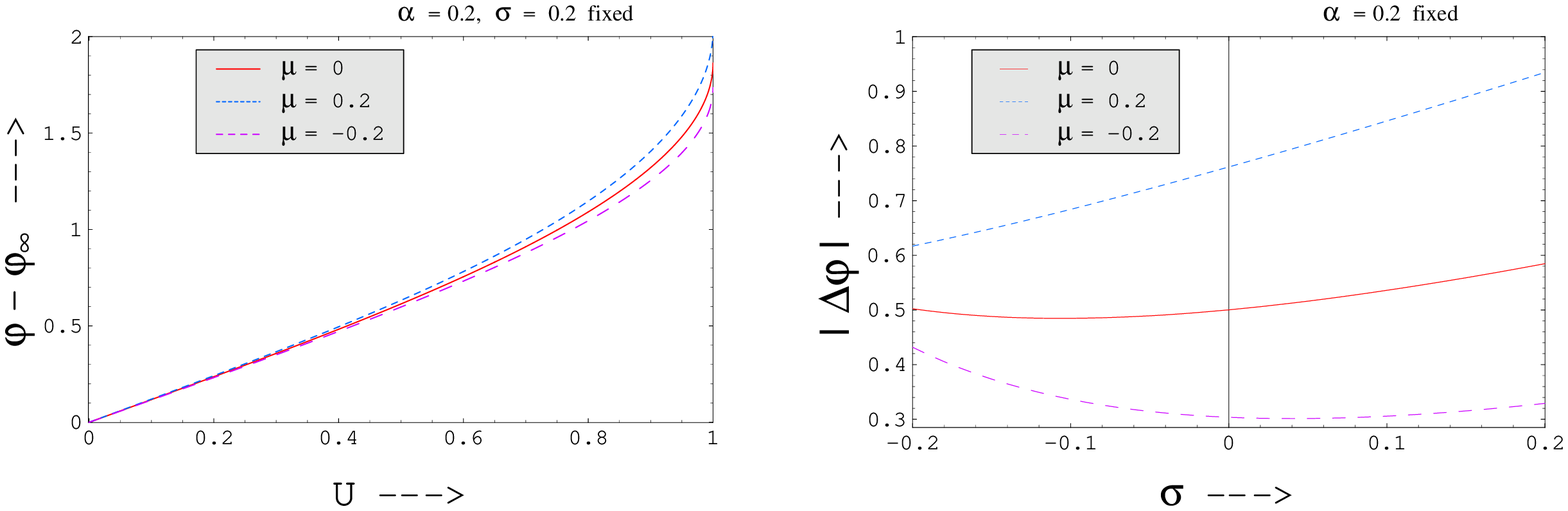}
{\small .}\hskip 1.9in (a) \hskip 3in (b)
\vskip .1in
\caption{ (a) Variation of $[\vp (r) - \vp_{\infty}]$ with $U = r_0/r$ for
characteristically chosen fixed values of the parameters $\alpha (= 0.2)$ and
$\sigma (= 0.2)$ and three values $(= 0, 0.2$ and $- 0.2)$ of the parameter
$\mu$. \hskip 1in
(b) plots of the magnitude of angle of bending $\Delta \vp$ as a function of $\sigma$ for a
fixed value of $\alpha (= 0.2)$ and three parametric values $(= 0, 0.2$ and $- 0.2)$
of $\mu$.}    \label{gr-spingb}
\end{figure}

\begin{figure}
\includegraphics[width=7in,height=3.25in]{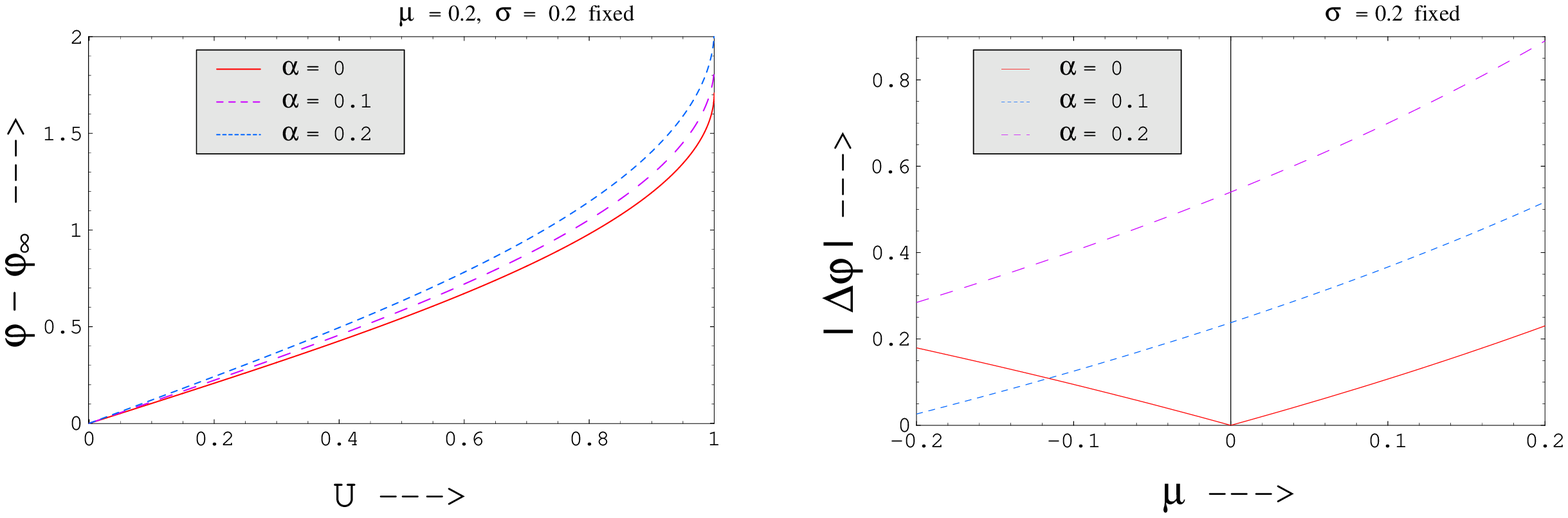}
{\small .}\hskip 1.9in (a) \hskip 3in (b)
\vskip .1in
\caption{ (a) Variation of $[\vp (r) - \vp_{\infty}]$ with $U = r_0/r$ for
characteristically chosen fixed values of the parameters $\mu (= 0.2)$ and
$\sigma (= 0.2)$ and three values $(= 0, 0.1$ and $0.2)$ of the parameter
$\alpha$. \hskip 1in
(b) plots of the magnitude of angle of bending $\Delta \vp$ as a function of $\mu$ for a
fixed value of $\sigma (= 0.2)$ and three parametric values $(= 0, 0.1$ and $0.2)$
of $\alpha$.}   \label{gr-spinga}
\vskip .1in
\end{figure}
It is easy to check that for $r_s = 0$, upto order $\sigma^2$, Eq.(\ref{gr-fspinorbit})
reduces exactly to Eq.(\ref{gr-kreomf}), the approximate analytic solution of
which is given in Eq.(\ref{gr-spinappsol}). For $r_s \neq 0$, however, in absence of such
analytic solution, we resort to numerical techniques to solve Eq.(\ref{gr-fspinorbit})
upto order $\sigma^2$. Now, in a general static spherisymmetric Einstein-KR spacetime,
one's prime interest is in the determination of effects of the particle's spin as
well as those of the KR field on standard astrophysical phenomena. For this purpose,
we need to study the trajectory deviation of spinning particles in the gravitational
field of non-compact objects whose mean radii are much larger than the Schwarzschild
radius $r_s$. In other words, the region $r \gg r_s$ is of particular relevance
in the context of KR and particle-spin effects.
Incidentally, both the metric coefficients $e^{\nu}$ and $e^{- \lambda}$
given in Eqs.(\ref{gr-krmetric}) are convergent for $r \gg r_s$, provided
the torsion (or, the KR field strength) is small, i.e., $|b|/r^2 \ll 1$,
in that domain. Since there is no experimental signature directly in favour
of torsion till now, it is reasonable to consider torsion to be weak, at
least, in the region $r \geq r_0$, with $r_0$ (the distance of closest approach
towards the center of force) much larger than the Schwarzschild radius. As such,
we consider the magnitude of the dimensionless KR measure $\mu = b/r_0^2$ to
be much smaller than unity. Dropping terms of order quadratic or more in $r_s/r$
and $b/r^2$ we can approximately write the solutions (\ref{gr-krmetric})
in terms of dimensionally scaled radial coordinate $U$ and the KR parameter
$\mu$ as
\be
e^{\nu (U)} ~=~ 1 ~-~ \alpha U ~;~~~~~~~~~
e^{- \lambda (U)} ~=~ 1 ~-~ \alpha U ~-~ \mu U^2.    \label{gr-krsolapp}
\ee
where $\alpha = r_s/r_0$.  For these solutions of the metric coefficients,
we obtain the numerical solutions of Eq.(\ref{gr-fspinorbit}) for various sets
of values of the dimensionless parameters $\alpha, \mu$ and $\sigma$. The
corresponding bending angle of trajectories ($\Delta \vp = 2 \vert\vp -
\vp_{\infty}\vert - \pi$) have been computed in various situations. The plots
of $(\vp - \vp_{\infty})$ as a function of $U$ as well as those of $\Delta \vp$
as a function of $\alpha$ or $\sigma$ or $\mu$ have been shown in the Figures
\ref{gr-spings} -- \ref{gr-spinga}.

Now to estimate the KR (or torson) measure $\mu$ (and also the spin
parameter $\sigma$) we follow the standard 
solar system analysis as in \cite{skssgss}. For bending of light near the sun, 
the impact parameter $r_0$ is roughly of the order of solar radius $R_s$ and
the parameter $\alpha$ is estimated to be
\be 
\alpha ~=~ \frac{r_s}{r_0} ~=~ \frac {2 G M_s}{c^2 R_s} 
\ee
where $M_s$ is the solar mass. Plugging the standard values for $M_s$ and $R_s$
\cite{wein,will} we find $\alpha$ to be extremely small ($\sim 5 \times 10^{-6}$).
The spin parameter $\sigma = K/(E I_g)$ can be estimated as follows:~ the value of
the spin angular momentum $K$ for photons is $\hbar$ and the energy $E = h c/\lambda$,
where $h$ is the Planck's constant, $\hbar = h/(2 \pi)$ and $c$ is the speed of
light. Therefore, $K/E = \lambda/(2 \pi c)$ which is numerically $\sim 10^{- 15}$ 
for visible
radiation with $\lambda \sim 5000 \AA$.  From Eq.(\ref{gr-Ig}), the constant $I_g$ 
can be given in terms of the parameters $r_0, \alpha$ and $\mu$ as
\be
I_g ~=~ \frac{r_0}{\sqrt{1 - \alpha}} ~-~ \frac K E \left(1 - \frac{3 \alpha} 2
\right) \sqrt{1 - \frac{\mu}{1 - \alpha}} .
\ee
With the small estimates of $\alpha$ and $K/E$ shown above, one can approximately 
write $I_g = r_0 \sim R_s$, which essentially implies an extremely small value of
the dimensionless spin parameter $\sigma ~(\sim 10^{- 23})$. Neglecting the small
estimates for $\alpha$ and $\sigma$, and integrating equations (\ref{gr-fspinorbit}) 
and (\ref{gr-fg}), we finally compute the leading contribution of the KR field on
the amount of bending of light trajectories as
\be
\Delta \vp |_{KR} ~=~ \frac {\pi \mu} 4 .
\ee
Using the error bars for the standard deflection of light measurements for the 
sun \cite{will} the parameter $\mu = b/r_0^2 \approx b/R_s^2$ turns out to be 
approximately of the order of $10^{- 6}$. 

\section{Conclusion}

We have clearly demonstrated the influence of a string inspired KR background on
the geodesics of spinning test particles. We have shown how the geodesic is modified
by various factors namely the KR field strength, spin of the particle and the gravitating
mass. The dependence of the geodesics on these factors have been exhibited graphically.
We hope that an accurate determination of the geodesic deviation of various cosmic particles in 
astrophysical experiments would be able to pinpoint the presence or absence of antisymmetric tensor field 
in the background spacetime leading to a much better understanding of the background spacetime geometry.
In addition, any evidence of presence of such a massless antisymmetric tensor field in the background 
may be looked upon as an indirect evidence of a string inspired low energy world.    

\vskip .2in
\noindent
{\bf {\Large Acknowledgment}}
\vskip .1in

DM and SS acknowledge the Council of Scientific and Industrial Research, Govt. of India for
providing financial support. SS also acknowledges Department of Atomic 
Energy, Government of India.

\end{document}